\documentclass[journal]{IEEEtran}

\usepackage{bm}
\usepackage{cite}
\usepackage{amsmath,amssymb,amsfonts}

\usepackage{booktabs}
\usepackage{tabularx}
\usepackage{threeparttable}

\usepackage[utf8]{inputenc}

\usepackage{hyperref}					
\hypersetup{colorlinks,
	linkcolor=blue,%
	anchorcolor=blue,
	citecolor=blue}

\usepackage{graphicx}
\usepackage{textcomp}
\usepackage{xcolor}
\usepackage{subfigure}

\begin{document}
	
	\title{\huge Clutter-Resilient ISAC for Low-Altitude Wireless Networks: \\A 5G Base Station-Compatible Protocol, \\ Waveform, and Prototype}
	
	\author{Jie~Wang, 
		Zhen~Du,~\IEEEmembership{Member,~IEEE}, Ying Wang, Weijie Yuan,~\IEEEmembership{Senior Member,~IEEE,} \\Fan Liu,~\IEEEmembership{Senior Member,~IEEE}, Xingdong Liang, and Yong Zeng,~\IEEEmembership{Fellow,~IEEE} 
		
		    \thanks{(\textit{Corresponding author: Zhen Du})}
			\thanks{Jie Wang, Zhen Du and Ying Wang are with the School of Electronic and Information Engineering, Nanjing University of Information Science and Technology, Nanjing 210044, China.}
			\thanks{Weijie Yuan is with the School of Automation and Intelligent Manufacturing, Southern University of Science and Technology, Shenzhen 518055, China.}
			\thanks{Fan Liu and Yong Zeng are with the National Mobile Communications Research Laboratory, Southeast University, Nanjing 210096, China.
				}
			\thanks{Xingdong Liang is with the National Key Laboratory of Microwave Imaging Technology, Aerospace Information Research Institute, Chinese Academy of Sciences, Beijing, China.}
	}

	\markboth{Journal of \LaTeX\ Class Files,~Vol.~xx, No.~xx, March~2026}%
	{Shell \MakeLowercase{\textit{et al.}}: Bare Demo of IEEEtran.cls for IEEE Journals}

	\maketitle

	\begin{abstract}
		Integrated sensing and communications (ISAC) has been envisioned as a promising solution to support emerging services in low-altitude wireless networks (LAWNs), where upgrading 5G ground base stations (GBS) toward new active sensing systems with wide coverage, low cost, high accuracy, and favorable spectrum compatibility, is strongly desired. However, such an evolution faces several critical challenges, particularly in the detection and tracking of weak and slow unmanned aerial vehicles (UAVs). These challenges include ISAC waveform design, clutter cancellation resilient to high clutter-to-noise ratios (CNRs), and efficient Doppler separation between UAVs and clutter. To that end, we summarize potential solutions and raise a comprehensive framework on implementing the 5G-advanced (5G-A) GBS. Outfield experiments demonstrate that the developed 5G-A GBS can effectively track weak and slow targets at distances exceeding $1$ kilometer, while incurring only a $1.2\%$ downlink rate loss relative to commercial 5G-A GBS. 
	\end{abstract}
	

	\IEEEpeerreviewmaketitle

	\section{Introduction}
	
	\subsection{Low-Altitude Wireless Networks: Vision and Requirements}	

	\IEEEPARstart{L}{ow}-altitude wireless networks (LAWNs) are envisioned as a key enabler for future smart cities and the emerging low-altitude economy \cite{wu2025low}. In such networks, a large number of heterogeneous platforms, such as small unmanned aerial vehicles (UAVs), electric vertical take-off and landing (eVTOL) vehicles, aerial sensors, and ground-air IoT devices, are expected to operate in the airspace below a few hundred meters. Typical application scenarios, such as parcel delivery, infrastructure inspection, emergency response, environmental monitoring, and traffic management, place stringent requirements not only on reliable wireless connectivity but also on timely awareness of the surrounding environment \cite{song2024overview,meng2023uav}.
	
	Compared with conventional terrestrial networks, LAWNs operate in a far more dynamic and complex propagation environment. The mobility of aerial platforms leads to rapidly time-varying channels, while the low-altitude space is characterized by rich scattering and strong reflections from buildings, vegetation, and terrain. These characteristics result in dynamic clutter, severe multipath effects, and intermittent blockage, which may significantly impair link reliability and security.
	
	To cope with these challenges, perception capabilities become an intrinsic requirement of LAWNs. Among various perception tasks, the detection and tracking of UAV activities are particularly critical for airspace management, collision avoidance, and interference control. Moreover, sensing environmental factors such as urban obstacles and rainfall patterns can further support link adaptation by anticipating shadowing and attenuation effects. While dedicated sensing devices, such as radars and lidars, are capable of providing such information, their deployment is often limited by high cost, restricted coverage, and the difficulty of supporting continuous wide-area operation.
	
	In this context, sensing enabled by ground base stations (GBS) emerges as a promising and scalable solution for LAWNs. Compared with onboard sensors carried by UAVs, GBS-based sensing benefits from wide-area coverage, stable power supply, and continuous operation. By leveraging existing communication infrastructure, GBSs can provide persistent awareness of low-altitude activities and environmental conditions, supporting functions such as UAV monitoring, obstacle awareness, and weather-related sensing. As a result, integrating sensing capabilities into terrestrial base stations is increasingly recognized as a key approach to meeting the evolving perception requirements of LAWNs.

	\begin{figure}[!t]
		\centering
		\includegraphics[width=3.5in]{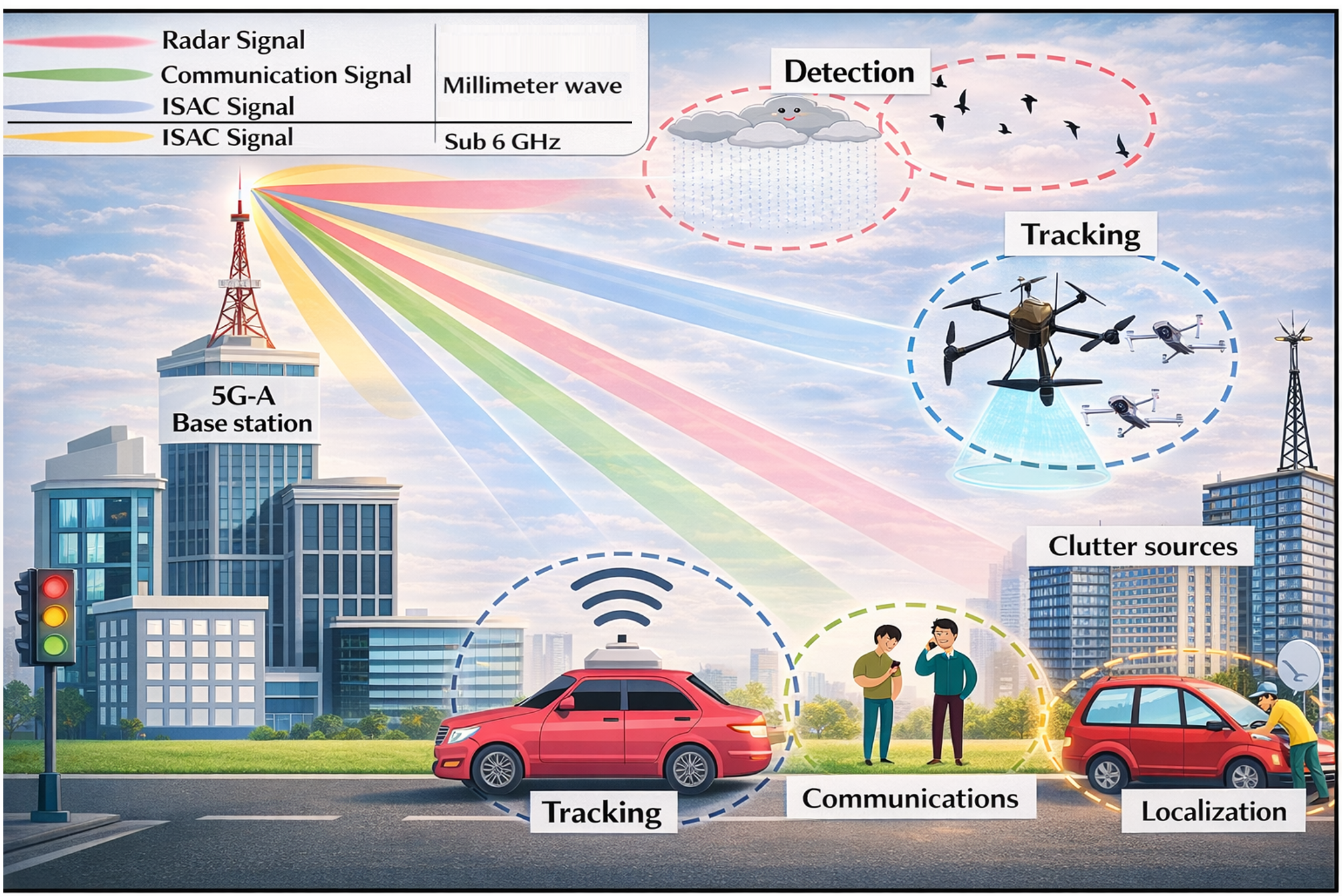}
		\caption{Overview of ISAC-empowered 5G-A services for LAWNs, where a single cellular infrastructure with two modes (wide beamwidth at sub-6 GHz and flexible beam management \cite{du2025toward} at millimeter wave), can simultaneously support downlink communications and environmental sensing, facilitating UAV perception and other representative low-altitude applications.}
		\label{fig0}
	\end{figure}

	\subsection{Why Does 5G-A GBS Need ISAC in LAWNs?}
	
	With the development of fifth-generation (5G) and emerging 5G-Advanced (5G-A) networks, cellular systems are gradually extending their functionality from pure connectivity provision to supporting environment-aware sensing. Although originally designed for communications, the physical-layer characteristics of GBS inherently support sensing functionalities. However, enabling sensing in operational cellular networks poses stringent constraints: downlink transmissions must remain communication-friendly, air-interface protocols must strictly comply with NR specifications, and sensing operations are required to coexist with ongoing traffic without incurring noticeable performance degradation. These constraints make it impractical to deploy sensing as an isolated functionality using dedicated waveforms or hardware.
	
	Under these circumstances, integrated sensing and communications (ISAC) offers a natural solution to these challenges by allowing the same waveform, hardware chain, and scheduling resources to be jointly utilized for both communication and sensing tasks \cite{du2025toward,wang2019first}. Rather than treating sensing as a separate end-goal, ISAC enables tight coupling between the two functions, thereby reducing hardware cost, improving spectrum efficiency, and supporting continuous wide-area perception within existing cellular infrastructure.
	
	Benefiting from dense deployment, continuous power supply, wide available bandwidth, and large-scale antenna arrays, 5G-A GBS are particularly well positioned to serve as perception nodes in LAWNs. By integrating sensing into the 5G-A GBS infrastructure, the network can support a range of perception functions, including UAV detection and tracking, obstacle awareness, and environmental monitoring, while simultaneously enhancing network operation through sensing-assisted beam management and resource scheduling \cite{du2025toward}, as illustrated in Fig.~\ref{fig0}. Consequently, 5G-A GBS establish a foundation for reliable real-time awareness in future LAWNs.

	\section{Technical Challenges}\label{sec2}
	Although 5G-A GBS may provide a promising pathway toward large-scale, cost-efficient perception in LAWNs, several fundamental challenges need to be addressed toward efficient sensing tasks.
	
	\subsection{Air-Interface and Waveform Layer: Embedding Sensing Signals under NR Protocol Constraints}
	In LAWNs, activating sensing functionality at GBS is fundamentally constrained by the communication-centric NR air interface. Unlike conventional radar systems, sensing signals at GBS cannot be freely designed or continuously transmitted. Instead, they must be embedded within standardized NR frame structures dominated by data traffic. As downlink resources are primarily allocated for communications, sensing is limited to fragmented time-frequency opportunities, while the fixed symbol structure further restricts waveform flexibility and precludes the direct use of radar-optimized signals. Consequently, sensing performance becomes highly dependent on scheduling and resource availability, in particular posing significant challenges for detecting weak and slow UAV targets in LAWNs. Therefore, the key challenge at the air-interface and waveform layer lies not in designing standalone optimal sensing waveforms, but in integrating sensing into NR-compliant transmissions without compromising existing orthogonal frequency division multiplexing (OFDM) waveform, which motivates the development of structured ISAC waveform paradigms under strict protocol constraints.
	
	\subsection{Physical Environment Layer: Ground Clutter and System Impairments}\label{sec2b}
\begin{figure}[!t]
		\centering 
		\subfigure[Without phase noise.]{
			\includegraphics[width=3.5in]{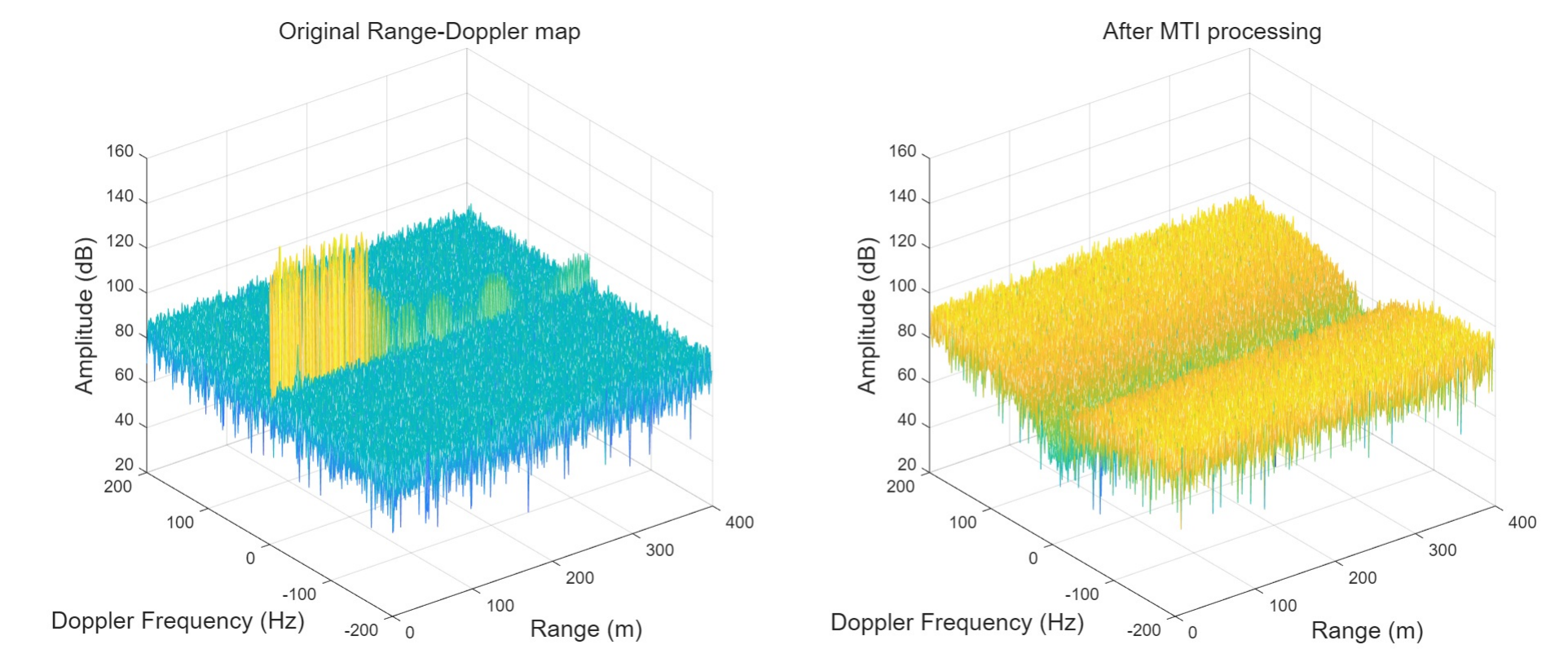}\label{f2a}}
		\subfigure[With phase noise.]{
			\includegraphics[width=3.5in]{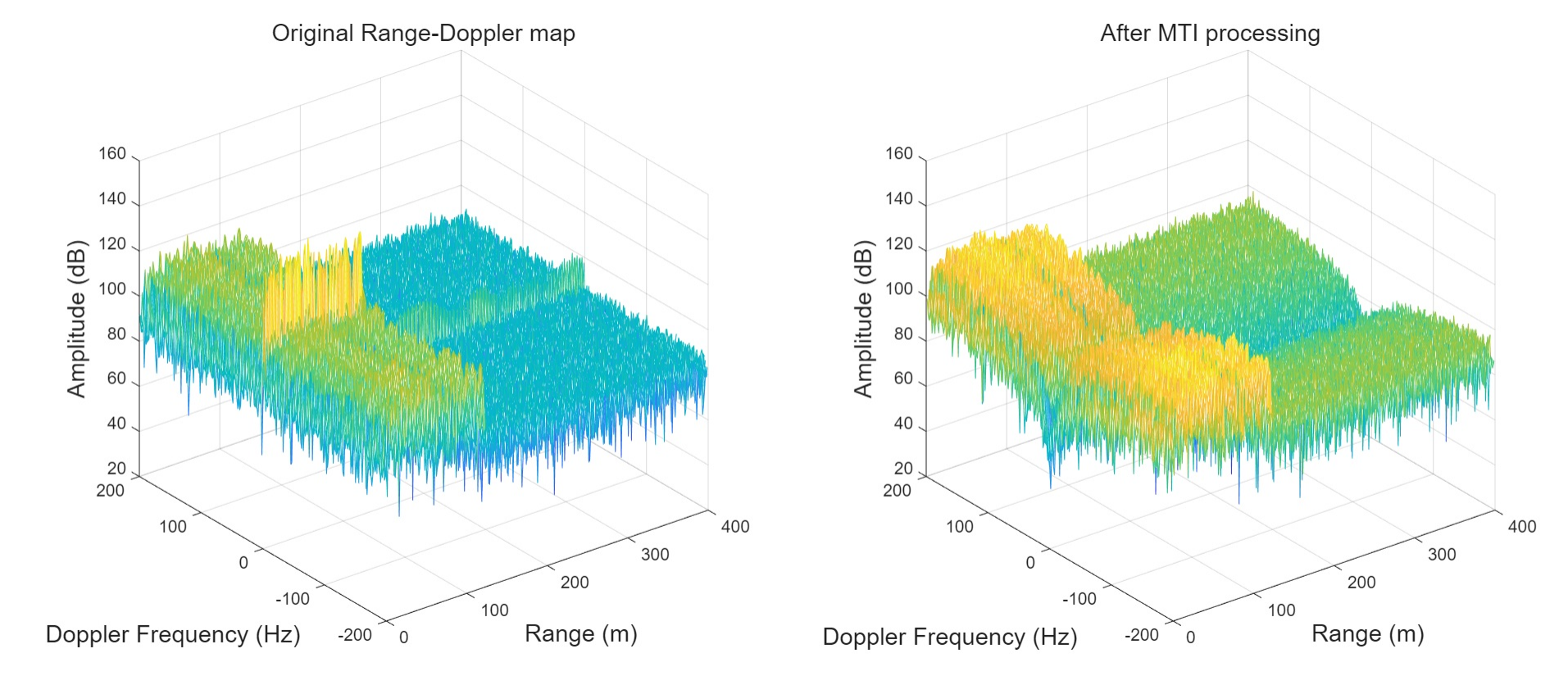}\label{f2b}}
		\caption{Clutter suppression through MTI.}\label{fig1}
	\end{figure}

	LAWNs are affected by reflections from buildings, vegetation, vehicles, and the ground itself, which generate dominant clutter and multipath interference
	Notably, LAWNs exhibit a high clutter-to-noise ratio (CNR), which fundamentally differs from traditional radar operations. Specifically, urban environments contain many static strong reflectors, such as wall-ground dihedrals, that dominate the received echo. In an ideal system, these scatterers would produce only a narrow zero-Doppler peak, which could be effectively removed by standard moving target indicator (MTI) \cite{richards2005fundamentals}, as illustrated in Fig.~\ref{f2a}. However, weak and slow UAVs near the zero-Doppler clutter may also be eliminated by MTI. Moreover, transceiver amplitude/phase jitter, shallow saturation, and other hardware imperfections cause these strong clutter scatterers to generate significant Doppler sidelobes, as depicted in Fig.~\ref{f2b}. These sidelobes which significantly elevate the pedestal floor across all Doppler bins in the same range cell, cannot be supressed by MTI, also impeding the detection of weak and slow UAVs. Consequently, these limitations motivate the need for more advanced clutter suppression techniques in LAWNs.
	
	\subsection{Signal Processing Layer: Reliable Detection and Tracking of Weak and Slow Targets}
	At the signal processing layer, challenges arising from the physical environment and air-interface constraints result in fundamental difficulties in target detection and tracking. Strong ground clutter, together with the limited sensing resources available within each NR frame, significantly distorts the statistical characteristics of the received signals. Moreover, weak and slow UAVs typically exhibit extremely small radar cross sections (RCS) (often $\leq 0.1 \ \mathrm{m}^2$), comparable to those of birds, and operate at low altitudes with low radial velocities. As a result, UAV echoes are frequently embedded within the main clutter band \cite{khawaja2025survey}, rendering reliable Doppler separation from clutter highly challenging. Consequently, false-alarm control becomes increasingly difficult, and classical detection assumptions, such as noise-limited test cells and well-separated targets \cite{richards2005fundamentals}, are often violated. These detection impairments further propagate to the tracking stage. Since tracking performance critically relies on the continuity and reliability of detection outputs (i.e., the measurement data), clutter-induced false alarms and missed detections can easily lead to track fragmentation or instability in LAWNs. This strong coupling between detection reliability and tracking robustness imposes stringent requirements on the signal processing chain for real-time low-altitude surveillance.

	\section{Potential Solutions}
	%

	To address the challenges identified in Sec.~\ref{sec2}, practical ISAC systems typically require a cross-layer design philosophy, where waveform configuration, sensing strategy, and receiver processing are jointly optimized.
	
	\subsection{Air-Interface and Waveform-Level Solutions}\label{sec3a}
	Under the stringent 5G-A air-interface constraints, ISAC sensing must be realized within the NR frame while coexisting with downlink communications, precluding the use of radar-centric waveforms. In this spirit, communication-centric waveforms \cite{du2025toward} can be adopted for sensing, which may be broadly categorized into three representative paradigms.
	
	\begin{itemize}
		\item \textbf{Pilot-based sensing:} This scheme exploits standardized NR reference signals, such as channel-state-information reference signals (CSI-RS) or sounding reference signals (SRS), for environmental perception. Owing to their full protocol compliance and predictable structure, pilot signals enable reliable matched filtering and low sidelobe levels. This approach is attractive for standard compliance and system simplicity, but its sensing performance is fundamentally constrained by the sparse pilot allocation in time and frequency.
		
		\item \textbf{Data-aided sensing:} By leveraging communication data symbols as sensing waveforms, the effective sensing aperture can be significantly enlarged. This approach can significantly improve SNR and resolutions, which is particularly beneficial for detecting weak and slow UAV targets. To counteract the non-ideal pulse compression properties caused by random constellations, advanced constellation design \cite{du2024reshaping}, or mismatch-aware processing techniques \cite{du2025probabilistic} may be employed.

		\item \textbf{Hybrid sensing:} This refers to schemes in which dedicated sensing symbols are sparsely embedded within NR frames \cite{wang2023waveform}. These symbols do not carry user data and can be flexibly designed using radar-friendly waveforms, such as chirps or phase-coded signals. By confining sensing to a small fraction of symbol periods, such hybrid schemes preserve communication performance while enabling controllable sensing capability. This paradigm is particularly appealing for practical 5G-A systems, where strict protocol compatibility and reliable sensing must be simultaneously ensured.
	\end{itemize}

	\subsection{Clutter Cancellation Solutions}
	To address the clutter-induced challenges, ISAC-enabled GBS can adopt environment-oriented strategies that aim to reshape the sensing background before target-focused processing. Rather than relying solely on sophisticated detectors, one may exploit structural properties of low-altitude environments and system operation to mitigate clutter at its source.
	
	\begin{itemize}
		\item \textbf{Doppler-guided separation of environment and targets:} A fundamental solution exploits the inherent Doppler contrast between static ground environments and moving aerial targets \cite{wang2023waveform}. By structuring sensing operations to emphasize Doppler-domain separation, clutter contributions concentrated around zero Doppler can be isolated and suppressed, effectively reducing the dynamic range of the sensing background. This strategy transforms clutter-dominated scenes into more tractable sensing conditions for slow UAV detection.
		
		\item \textbf{Environment-aware background estimation and removal:} In quasi-static low-altitude deployments, dominant clutter components remain stable over long durations. This enables the construction of environment-aware background representations, such as static reflection profiles or long-term averaged range-Doppler maps. By explicitly estimating and suppressing these environment-induced components, target echoes can be effectively preserved from clutter corruption. For instance, a channel knowledge map (CKM) \cite{zeng2024tutorial} can be leveraged to store offline-estimated clutter angles, thereby facilitating online clutter mitigation prior to target detection and parameter estimation.
		
		\item \textbf{Clutter-resilient system operation under non-idealities:} Beyond ideal propagation assumptions, strong static scatterers combined with hardware non-idealities can introduce elevated Doppler sidelobes and background leakage. To counteract this effect, clutter-resilient system operation strategies can be adopted, including controlled sensing waveform scheduling, dynamic range management, and mitigation of saturation-induced distortions, etc. These system-level measures prevent excessive clutter-induced background elevation and maintain reliable sensing performance.
	\end{itemize}
	

	\subsection{Signal Processing Solutions for Weak and Slow Targets}
	Detecting weak and slow UAVs in LAWNs suffers from several challenges. Even after clutter cancellation, the residual background is often non-homogeneous, and the target echoes may be buried beneath strong multipath or residual sidelobes. Existing solutions can be broadly classified into three complementary approaches.
	
	\begin{itemize}
		\item \textbf{Adaptive thresholding and CFAR-based detection:}	
		A fundamental approach relies on adaptive detection schemes that adjust thresholds according to local noise or clutter statistics \cite{richards2005fundamentals}. Constant false alarm rate (CFAR) detectors, including cell-averaging, order-statistic, and trimmed-mean variants, can be extended to handle spatially or temporally varying backgrounds. By dynamically estimating the local noise floor, these methods maintain reliable detection performance even in regions with residual clutter sidelobes or non-stationary interference.
		
		\item \textbf{Model-based integration and coherent processing:}
		To enhance sensitivity for weak and slow targets, coherent integration across multiple symbols, subcarriers, or antenna channels can be exploited \cite{liu2025uncovering}. Model-based approaches can leverage prior knowledge of target motion or expected Doppler characteristics to accumulate signal energy constructively. This includes matched filtering, space-time adaptive processing (STAP), and multi-antenna beamforming schemes that increase SNR while suppressing residual clutter and interference.
		
		\item \textbf{Learning-assisted and data-driven detection:}
		Complementary to model-based methods, learning-assisted or data-driven techniques can adaptively optimize detection in complex residual backgrounds \cite{Shi2025}, which is particularly beneficial to LAWNs. For instance, neural networks, support vector machines (SVM), or Bayesian inference frameworks can be trained to distinguish target component from non-Gaussian residual clutter. These methods are particularly useful in highly dynamic scenarios, providing robustness beyond classical model assumptions.
	\end{itemize}
	
	The above signal processing strategies collectively address the challenge of detecting weak and slow UAVs in ground clutter. In practice, one may exploit hybrid implementations by combining adaptive thresholding, coherent integration, and learning-assisted detection to achieve high reliability in clutter-dominated LAWNs.

	\begin{figure*}[!t]
		\centering
		\includegraphics[width=6.95in]{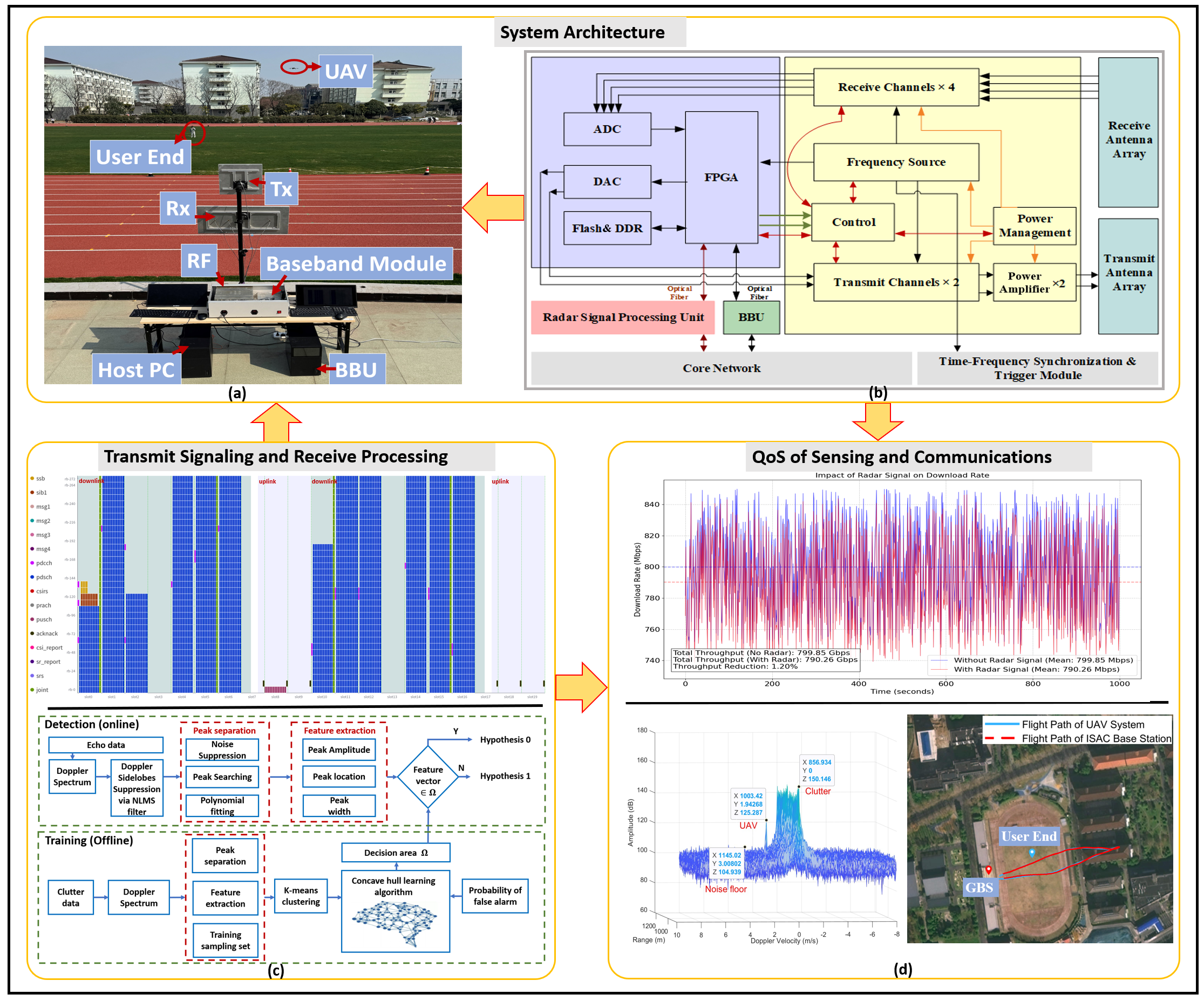}
		\caption{A systematic framework implementation of ISAC-empowered 5G-A GBS and results.}
		\label{f5}
	\end{figure*}

	%
	%

	\section{Prototype Implementation and Experiments}
	This section demonstrates the practical efficacy of our developed 5G-A GBS prototype. Overall, the mono-static ISAC prototype complies with standard NR frame structures at sub-6 GHz and does not modify the communication protocol, thereby ensuring full compatibility with current 5G-A deployments.

	%
	
	\subsection{A Systematic Framework}\label{sec4a}
	\subsubsection{Prototype Architecture}
	The prototype is illustrated in Fig.~\ref{f5}(a) and (b). First, to enable the sensing functionality at the GBS, target echoes from the downlink signals must be captured during the downlink time slots. To achieve this, we adopt a spatially separated transmit and receive architecture for the ISAC system, ensuring rigorous isolation against direct-path self-interference. Compared to existing commercial 5G GBSs, this configuration requires only four additional antenna pairs. Second, the devised 5G-A GBS primarily consists of a baseband processing unit (BBU), a remote radio unit (RRU), and antennas, which is similar to commercial 5G GBS. Specifically, the BBU handles the modulation and demodulation of ISAC signals, radar signal processing, and target detection, etc. During the transmission link, the RRU receives baseband signals from the BBU and processes them through digital-to-analog conversion, up-conversion, filtering, and amplification before transmitting them via the antennas. In the reception link, the antenna receives signals and delivers them to the RRU, which then processes the signals through low-noise amplification, filtering, down-conversion, and analog-to-digital conversion before delivering them to the BBU. Overall, this ISAC framework necessitates the continuously operational reception channels of RRU, allowing the receiving antennas to simultaneously capture sensing echoes and uplink communication signals with a mode of time division duplex (TDD).

	\subsubsection{ISAC Configuration}
	Motivated by the hybrid sensing paradigm discussed in Sec.~\ref{sec3a}, we implement a 5G-A ISAC GBS where a small number of chirp-based sensing symbols are periodically inserted into the downlink frame. Compared with current 5G NR protocol, our framework merely optimizes the physical layer, which is rather limited and will not deteriorate the access or control of wireless communication terminals. To further achieve sensing task with minimal resources, it is necessary to optimize the design of the physical layer's resource mapping and demapping modules in the user plane.
	In particular, although the last symbol of each downlink slot is nominally reserved for the PDCCH, it is frequently underutilized in practical deployments. Leveraging this opportunity, we configure these symbols as sensing symbols carrying chirp spectra on subcarriers, thereby enabling sensing without compromising communication scheduling or inducing sensing-communications interference. Notably, the sensing symbols should be carefully configured. For example, as depicted in Fig.~\ref{f5}(c), the slot configuration of the 5-ms composite frame contains successive 7 downlink slots, followed by a special slot and 2 uplink slots. Consequently, to ensure a uniform slow-time sampling, we may select slot 0 and slot 5 of each downlink frame period. As such, the pulse repetition interval (PRI) is 5 slots, corresponding to 0.5 ms. In this manner, the sensing overhead occupation is approximately $1\%$.

	\subsubsection{ISAC Signal Processing}
\begin{figure}[!t]
		\centering 
		\subfigure[2D-CFAR detection without MTI.]{
			\includegraphics[width=1.65in]{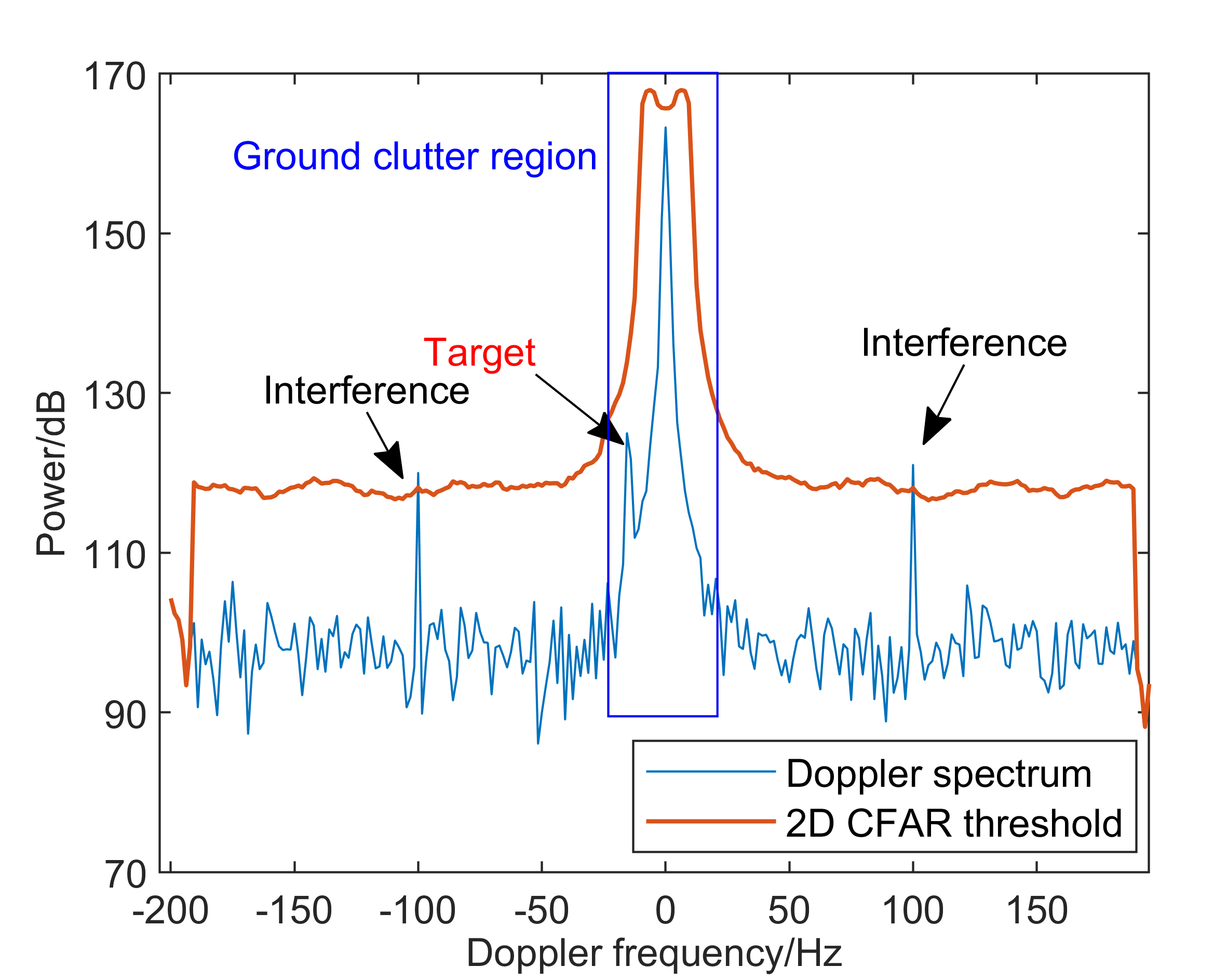}\label{f4a}}
		\subfigure[2D-CFAR detection with MTI.]{
			\includegraphics[width=1.65in]{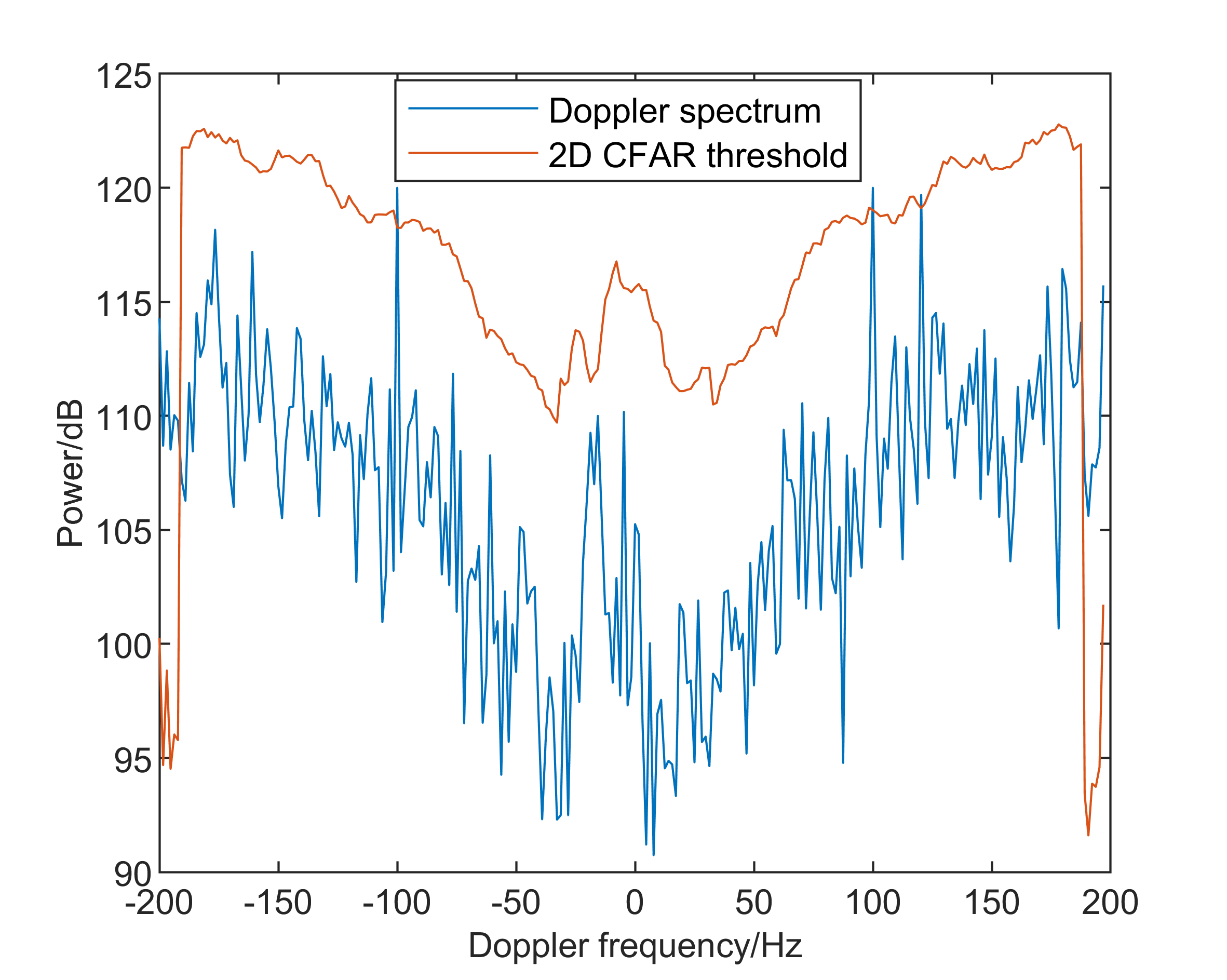}\label{f4b}}
		\subfigure[Proposed CFPS-CFAR detection: peak separation.]{
			\includegraphics[width=3.25in]{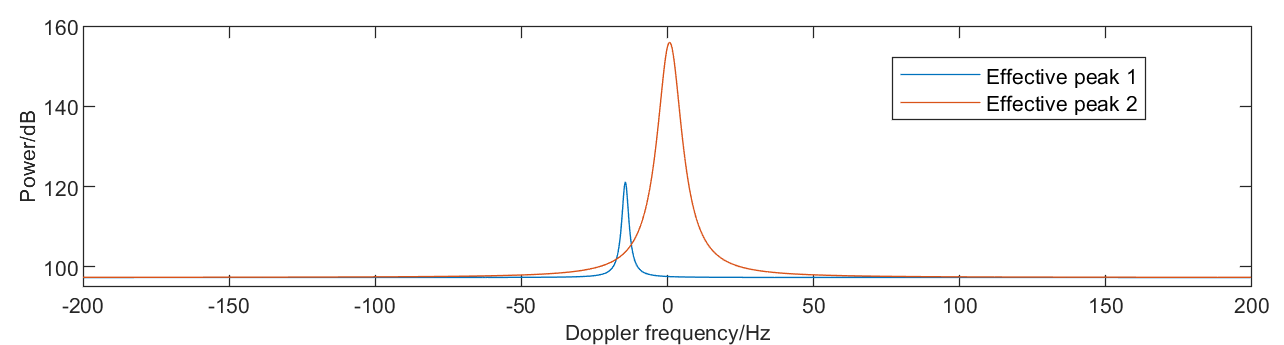}\label{f4c}}
		\caption{Detection of the weak and slow UAV target.}\label{fig4}
	\end{figure}

	The sensing receive processing includes clutter cancellation, target detection, and tracking.
	\begin{itemize}
		\item \textbf{Clutter cancellation:} To address the pronounced Doppler sidelobes of near-range clutter scatterers induced by phase noise and shallow saturation, a classic normalized least-mean-square (NLMS) algorithm is firstly employed before detection. By estimating the steady-state NLMS filter coefficients, Doppler sidelobes can be effectively suppressed. However, after NLMS processing, there may still exhibit a high CNR near zero-Doppler range cells, which significantly degrades target detection performance within the main clutter band. 
		
		\item \textbf{Target detection:} After the NLMS operation, the weak and slow UAV target remains undetected by the 2D-CFAR detector, as it is located in the vicinity of the zero-Doppler clutter region, as illustrated in Fig.~\ref{f4a}. One may naturally apply MTI following NLMS filtering to further suppress the clutter. However, this operation also significantly attenuates the target signal, rendering the 2D-CFAR detector ineffective as well, as shown in Fig.~\ref{f4b}. To overcome these limitations, a feature-based intelligent detection framework is devised, which reformulates the classical target detection problem as a classification task in a feature space. In particular, noting that clutter samples are overwhelmingly more abundant than target samples, conventional supervised learning methods become less effective. Leveraging this imbalance, low-altitude target detection under strong ground clutter is modeled as an anomaly detection problem, where clutter echoes are treated as normal samples and target-containing echoes as anomalies. Accordingly, the proposed detector is trained only on clutter samples, without requiring target-labeled data. As specified in Fig.~\ref{f5}(c), in the detection branch, a curve-fitting and peak-separation CFAR (CFPS-CFAR) detector is developed. The echoes are first transformed into the Doppler domain, followed by noise suppression and derivative-based peak searching. Polynomial fitting is then applied to separate target and clutter components, from which distinguishable features such as peak amplitude, location, and width can be extracted, as depicted in Fig.~\ref{f4c}. In the training branch, a fast concave hull learning algorithm \cite{Shi2025} is employed to generate the decision region of clutter samples in the feature space, given a false alarm rate. Finally, exploiting the zero-Doppler characteristic of stationary clutter, the peak location is used as the test statistic to enable binary detection.
		
		\item \textbf{Target tracking:} After target detection, tracking is carried out within a standard Kalman filtering framework. Specifically, the target motion state is assumed to remain constant within each slot and can follow multiple motion models, such as constant-velocity, constant-acceleration/deceleration, or coordinated-turn models. Based on this, an interacting multiple-model extended Kalman filter (IMM-EKF) \cite{meng2023vehicular} is employed to fuse the measurements and state information, thereby enabling stable real-time estimation of the target trajectory.
	\end{itemize}
	
	Finally, since the embedding of sensing symbols does not affect the communication symbols in the NR frame, the communication signal processing still follows the standard NR procedures and is therefore omitted in this article.
	

	\subsection{Experiments of UAVs' Perception}
	The outfield scene is shown in Fig. \ref{f5}(a), where the 5G-A GBS can simultaneously support communications to a user end and sensing toward a weak and slow UAV (marque: DJI Mavic 3). System parameters are listed as follows:
	\begin{itemize}
		\item Frequency band ranges from 1.99872~GHz to 5.04~GHz; 
		\item Signaling bandwidth is $100$ MHz; 
		\item Subcarrier bandwidth is $30$ KHz, coinciding with the frame structure of a $5$ ms single cycle in NR;
		\item Accumulated pulse number is 256;
		\item Transmit gain ranges from 0~dB to 31~dB; 		
		\item Maximum transmit power of a single carrier is 40~W;
		\item Antenna gain is 12~dB; 	
		\item 4 Tx and 4 Rx are configured, where the isolation between Tx channels is larger than 50~dB, the isolation between Rx channels is larger than 90~dB, and the isolation between Tx and Rx channels is larger than 90~dB.
	\end{itemize}
	
	The experimental range-velocity map at the distance of 1 kilometer is also depicted in Fig.~\ref{f5}(d), where the output SNR ($\mathrm{SNR_{out}}$) after matched filtering and clutter cancellation is almost 20~dB.
	To theoretically evaluate $\mathrm{SNR_{out}}$, we exploit the celebrated radar equation \cite{richards2005fundamentals}
	\begin{align}
		\mathrm{SNR_{out}} = \frac{P_tG^2\lambda^2\sigma N_\mathrm{sym}(T\cdot \eta)}{(4\pi)^3R^4_\mathrm{max}kT_0L_\mathrm{loss}},
	\end{align}
	where the transmit power ($P_t$) is approximately 40~dBm\footnote{In order to avoid the non-linear distortion of power amplifier, the back-off power 40~dBm is exploited.}, the antenna gain ($G$) is 12 dB, the carrier frequency is 3.747~GHz (leading to the wavelength ($\lambda$) as $80.064$ mm), the UAV's RCS ($\sigma$) is $0.1$~$\mathrm{m}^2$, accumulated sensing symbol number ($N_\mathrm{sym}$) is $256$, the symbol duration ($T$) is $33.33~\mu s$, the radar frequency duty ratio ($\eta$) is $50\%$, the maximum detection distance ($R_\mathrm{max}$) is $1$ km, $k$ is Boltzmann's constant, the standard temperature ($T_0$) is 290 K, and the noise coefficient ($L_\mathrm{loss}$) is 8~dB. Therefore, the calculated $\mathrm{SNR_{out}}$ is approximately 21~dB, which coincides with the experimental result.
	Although an output SNR of 20~dB still leaves a considerable margin with respect to the maximum detectable range, a further reduction in SNR would lead to a pronounced degradation in angle estimation accuracy, eventually preventing reliable real-time target tracking. As a consequence, the maximum detectable distance exceeds 1 kilometer. 
	
	The real-time UAV tracking trajectory is also illustrated in Fig.~\ref{f5}(d), together with the reference flight trajectory provided by the UAV onboard system for comparison. A close agreement between the two trajectories can be observed, which validates the power of our prototype.
	
	Finally, the communication performance is evaluated in Fig.~\ref{f5}(d) by measuring the real-time downlink rate. With the chirp symbols embedded as described in Sec.~\ref{sec4a}, the average downlink rate over 1000 seconds reaches approximately 791.18~Mbps, in contrast to 800.79~Mbps of the communication-only benchmark. As such, we highlight the fact that, our constructed framework of ISAC-empowered 5G-A GBS can effectively track low-altitude, small, and slow UAVs exceeding 1 kilometer, incurring only a $1.2\%$ throughput reduction\footnote{This is superior to current commercial 5G-A GBS. For example, it has been reported that Huawei conducted the world's first verification of 5G-A ISAC capability at the Huairou Outfield in Beijing \cite{huawei}. Overall, the integrated 5G ISAC sensor, utilizing the 3GPP 5G signal in the millimeter-wave band, is capable of perceiving vehicles and people at a distance of over 500~m with the proportion of perception resources up to 15\%.}.


	\section{Open Issues and Opportunities}
	\subsection{Multi-Modal Fusion Perception}
	In addition to ISAC-based scheme, other sensory modalities such as video and acoustics perception, have also been broadly employed in LAWNs. Generally, wireless sensing (including radar and 5G-A GBS) can provide a superior ranging ability, voice behaves better in angle measurement ability, while video is more qualified to classify multiple objects. In contrast, video is more susceptible to meteorological conditions, voice is sensitive to noise, whereas wireless sensing is more robust against these adverse factors. Consequently, the effective fusion of multi-modal perception remains as an attractive solution to enhance  surveillance behaviors of low-altitude UAVs.
	
	\subsection{Meteorological Information Perception With 5G-A GBS}
	The effect of atmospheric impairments on radio transmission has been widely reported, while utilizing 5G communication signals reciprocally to infer meteorological information is rarely reported. In practice, 5G-A GBS may potentially obtain meteorological information through its perception technologies, such as wireless sensing and millimeter-wave communications. For example, when accessed communication users are fewer at night, 5G-A GBS may adjust their elevation angles to obtain abundant data on the structure and movement of clouds, while the attenuation and reflection of communication signals can be used to infer atmospheric humidity.
	
	\subsection{Multi-GBS Cooperative UAV Sensing}
	Beyond single-station sensing, cooperative UAV perception enabled by multi-GBS offers a promising means to improve sensing reliability and coverage in LAWNs. By jointly exploiting spatially distributed observations, multi-GBS cooperation can mitigate blockage and unfavorable geometries, leading to enhanced detection robustness and localization accuracy for low-altitude UAVs.
	Despite these advantages, multi-GBS cooperative sensing poses challenges in inter-station synchronization, information sharing, and scalable fusion under heterogeneous sensing conditions.

	\section{Conclusion}
	In this review article, we present a systematic framework for joint wireless communications and clutter-resilient sensing in LAWNs, including protocol design, waveform configuration, and 5G-A GBS prototype implementation. We begin by identifying key technical challenges in developing ISAC-empowered 5G-A GBS in LAWNs and outlining their potential solutions. Building on these insights, a complete prototype framework is developed, detailing the protocol stack architecture, frame structure, and signal processing schemes, among other components. Sensing and communication performance are further evaluated through outfield experiments, demonstrating the effectiveness of the proposed scheme with a maximum detectable distance exceeding $1$ kilometer, while incurring only a 1.2\% downlink rate loss.

	\ifCLASSOPTIONcaptionsoff
	\newpage
	\fi
	
	\bibliographystyle{IEEEtran}
	\bibliography{reference}

@article{khawaja2025survey,
  title={A survey on detection, classification, and tracking of {UAV}s using radar and communications systems},
  author={Khawaja, Wahab and Ezuma, Martins and Semkin, Vasilii and Erden, Fatih and Ozdemir, Ozgur and Guvenc, Ismail},
  journal={IEEE Communications Surveys \& Tutorials},
  year={2025},
  publisher={IEEE}
}

@article{zeng2024tutorial,
  title={A tutorial on environment-aware communications via channel knowledge map for 6{G}},
  author={Zeng, Yong and Chen, Junting and Xu, Jie and others},
  journal={IEEE Communications Surveys \& Tutorials},
  volume={26},
  number={3},
  pages={1478--1519},
  year={2024},
  publisher={IEEE}
}

@misc{huawei,
  author= {Huawei},
  title={Huawei successfully completes 5{G}-{A}dvanced synesthesia verification},
  howpublished = {\url{https://www.gizchina.com/huawei/huawei-is-the-first-to-complete-5g-advanced-synesthesia-verification}},
  note = {Accessed: Dec. 31, 2021}
}

@article{meng2023vehicular,
  title={Vehicular connectivity on complex trajectories: Roadway-geometry aware {ISAC} beam-tracking},
  author={Meng, Xiao and Liu, Fan and Masouros, Christos and others},
  journal={IEEE Transactions on Wireless Communications},
  volume={22},
  number={11},
  pages={7408--7423},
  year={2023},
  publisher={IEEE}
}

@article{wang2019first,
  title={First demonstration of joint wireless communication and high-resolution {SAR} imaging using airborne {MIMO} radar system},
  author={Wang, Jie and Liang, Xing-Dong and Chen, Long-Yong and others},
  journal={IEEE Transactions on Geoscience and Remote Sensing},
  volume={57},
  number={9},
  pages={6619--6632},
  year={2019},
  publisher={IEEE}
}

@article{Shi2025,
  title={False-Alarm-Controllable Detection of Marine Small Targets via Improved Concave Hull Classifier},
  author={Shi, Sainan and Wang, Jiajun and Wang, Jie and others},
  journal={Remote Sensing},
  volume={17},
  number={11},
  pages={1808},
  year={2025},
}

@article{liu2025uncovering,
  title={Uncovering the iceberg in the sea: Fundamentals of pulse shaping and modulation design for random {ISAC} signals},
  author={Liu, Fan and Xiong, Yifeng and Lu, Shihang and Li, Shuangyang and Yuan, Weijie and Masouros, Christos and Jin, Shi and Caire, Giuseppe},
  journal={IEEE Transactions on Signal Processing},
  year={2025},
  publisher={IEEE}
}

@article{du2025toward,
  title={Toward {ISAC}-Empowered Vehicular Networks: Framework, Advances, and Opportunities},
  author={Du, Zhen and Liu, Fan and Li, Yunxin and others},
  journal={IEEE Wireless Communications},
  volume={32},
  number={2},
  pages={222--229},
  year={2025},
  publisher={IEEE}
}

@article{wu2025low,
  title={Low-altitude wireless networks: A survey},
  author={Wu, Jun and Yang, Yaoqi and Yuan, Weijie and others},
  journal={arXiv:2509.11607},
  year={2025}
}

@book{richards2005fundamentals,
  title={Fundamentals of radar signal processing},
  author={Richards, Mark A},
  volume={1},
  year={2005},
  publisher={Mcgraw-hill New York}
}

@article{du2025probabilistic,
  title={Probabilistic Constellation Shaping for {OFDM ISAC} Signals Under Temporal-Frequency Filtering},
  author={Du, Zhen and Xu, Jingjing and Xiong, Yifeng and others},
  journal={IEEE Transactions on Wireless Communications},
  year={2026}
}

@article{du2024reshaping,
  title={Reshaping the {ISAC} tradeoff under {OFDM} signaling: A probabilistic constellation shaping approach},
  author={Du, Zhen and Liu, Fan and Xiong, Yifeng and others},
  journal={IEEE Transactions on Signal Processing},
  year={2024},
  publisher={IEEE}
}

@article{song2024overview,
  title={An overview of cellular {ISAC} for low-altitude {UAV}: New opportunities and challenges},
  author={Song, Yuxuan and Zeng, Yong and Yang, Yuhang and others},
  journal={IEEE Communications Magazine},
  year={2025},
  publisher={IEEE}
}

@article{wang2023waveform,
  title={Waveform designs for joint wireless communication and radar sensing: Pitfalls and opportunities},
  author={Wang, Jie and Liang, Xing-Dong and Chen, Long-Yong and others},
  journal={IEEE Internet of Things Journal},
  volume={10},
  number={17},
  pages={15252--15265},
  year={2023},
  publisher={IEEE}
}

@article{meng2023uav,
  title={{UAV}-enabled integrated sensing and communication: Opportunities and challenges},
  author={Meng, Kaitao and Wu, Qingqing and Xu, Jie and others},
  journal={IEEE Wireless Communications},
  year={2023},
  publisher={IEEE}
}
	
	\vspace{5mm}
	\noindent \textbf{Jie Wang} (wangjie110@nuist.edu.cn) is an Associate Professor with the School of Electronic and Information Engineering, Nanjing University of Information Science and Technology, Nanjing, China. 
	
	\vspace{5mm}
	\noindent \textbf{Zhen Du} [M] (duzhen@nuist.edu.cn) is an Associate Professor with the School of Electronic and Information Engineering, Nanjing University of Information Science and Technology, Nanjing, China.
	
	\vspace{3mm}
	\noindent \textbf{Ying Wang} (202312490171@nuist.edu.cn) is a master student with the School of Electronic and Information Engineering, Nanjing University of Information Science and Technology, Nanjing, China. 
	
	\vspace{3mm}
	\noindent \textbf{Weijie Yuan} [SM] (yuanwj@sustech.edu.cn) is an Assistant Professor with the School of Automation and Intelligent Manufacturing, Southern University of Science and Technology, Shenzhen, China. 
	
	\vspace{3mm}
	\noindent \textbf{Fan Liu} [SM] (fan.liu@seu.edu.cn) is a Professor at the National Mobile Communications Research Laboratory, Southeast University, Nanjing, China.
	
	\vspace{3mm}
	\noindent \textbf{Xingdong Liang} (xdliang@mail.ie.an.cn) is a Professor with the National Key Laboratory of Microwave Imaging Technology, Aerospace Information Research Institute, Chinese Academy of Sciences.
	
	\vspace{3mm}
	\noindent \textbf{Yong Zeng} [F] (yong\_zeng@seu.edu.cn) is a Professor at the National Mobile Communications Research Laboratory, Southeast University, Nanjing, China.
	
\end{document}